\documentclass[a4paper,11pt]{article}
\usepackage{pos}

\usepackage{amsmath}
\usepackage{amsfonts}
\usepackage{braket}
\usepackage{slashed}
\usepackage{simplewick}

\usepackage{siunitx}

\usepackage{graphicx}
\usepackage{float}

\usepackage{listings}

\usepackage{tikz}
\usetikzlibrary{calc}
\usetikzlibrary{patterns}
\usetikzlibrary{patterns}
\usetikzlibrary{positioning,decorations.pathmorphing,decorations.markings}

\usepackage{feynmp-auto}
\usepackage{multirow}

\usepackage{xcolor}
\usepackage{overpic}

\title{Lattice QCD Constraints on the Fourth Mellin Moment of the Pion Light Cone Distribution Amplitude using the HOPE method}
\ShortTitle{Fourth Mellin Moment of the Pion LCDA using the HOPE Method}

\manuallySeparateAuthors
\author[a,b]{William Detmold,}
\author[a,c]{ Anthony V. Grebe,}
\author[d]{ Issaku Kanamori,}
\author[e,f]{ C.-J. David Lin,}
\author*[g]{ Robert J. Perry}
\author[h]{ and Yong Zhao}
\author{ for the HOPE Collaboration}

\affiliation[a]{Center for Theoretical Physics, Massachusetts Institute
  of Technology,
Cambridge, MA 02139, USA}
\affiliation[b]{The NSF AI Institute for Artificial Intelligence and Fundamental Interactions
}
\affiliation[c]{Theoretical Physics Department, Fermi National Accelerator Laboratory,
P.O. Box 500, Batavia, IL 60410, USA}
\affiliation[d]{RIKEN Center for Computational Science, Kobe 650-0047, Japan}
\affiliation[e]{Institute of Physics, National Yang Ming Chiao Tung University, 1001 Ta-Hsueh Road, Hsinchu 30010, Taiwan}
\affiliation[f]{Centre for High Energy Physics, Chung-Yuan Christian University, Chung-Li, 32032, Taiwan}
\affiliation[g]{Departament de F\'isica Qu\`antica i Astrof\'isica and Institut de Ci\`encies del Cosmos, Universitat de Barcelona, Mart\'i Franqu\`es 1, E08028, Spain}
\affiliation[h]{Physics Division, Argonne National Laboratory, Lemont, IL 60439, USA}

\emailAdd{wdetmold@mit.edu}
\emailAdd{agrebe@mit.edu}
\emailAdd{kanamori-i@riken.jp}
\emailAdd{dlin@nycu.edu.tw}
\emailAdd{perryrobertjames@gmail.com}
\emailAdd{yong.zhao@anl.gov}

\abstract{The light-cone distribution amplitude (LCDA) of the pion contains information about the parton momentum carried by the quarks and is an important theoretical input for various predictions of exclusive processes at high energy, including the pion electromagnetic form factor. Progress towards constraining the fourth Mellin moment of the LCDA using the heavy-quark operator product expansion (HOPE)
method is presented.}

\FullConference{The 40th International Symposium on Lattice Field Theory (Lattice 2023)\\
July 31st - August 4th, 2023\\
Fermi National Accelerator Laboratory\\}

\begin{document}
\maketitle

\section{Introduction}
The pion light-cone distribution amplitude (LCDA) is a non-perturbative quantity required for the description of a range of exclusive processes in high-energy quantum chromodynamics (QCD)~\cite{Lepage:1980fj}. It is denoted $\phi_\pi(\xi,\mu^2)$ and is defined via the matrix element of the transition between the vacuum and a charged pion state,
\begin{equation}
\braket{0|\overline{\psi}_d(z)\gamma_\mu \gamma_5 \mathcal{W}[z,-z]\psi_u(z)|\pi^+(\mathbf{p})}=if_\pi p_\mu\int_{-1}^{1}d\xi\, e^{-i\xi p\cdot z}\phi_\pi(\xi,\mu^2)\,,
\end{equation}
where $\mathcal{W}[z,-z]$ is a light-like ($z^2=0$) Wilson line connecting $-z$ and $z$ and $\mu$ is the renormalization scale. In the above equation, $f_\pi$ is the pion decay constant and $p^\mu$ is the four-momentum of the pion. In the light-cone gauge, the pion LCDA can be interpreted as the probability amplitude to convert the pion into a state of a quark and an antiquark carrying momentum fractions $(1 + \xi)/2$ and $(1 - \xi)/2$, respectively. 

Such non-perturbative quantities are natural targets of lattice QCD (LQCD) calculations. However, many of the matrix elements appearing in hadron structure calculations contain non-local operators defined with light-like separations. Matrix elements of such operators cannot be directly accessed in a Euclidean field theory. As a result, in the last two
decades a number of alternative strategies for extracting information about these non-perturbative matrix elements
using LQCD have been proposed~\cite{Aglietti:1998ur,Liu:1999ak,Detmold:2005gg,Braun:2007wv,Davoudi:2012ya,Ji:2013dva,Chambers:2017dov,Radyushkin:2017cyf,Ma:2017pxb}. This work follows the approach suggested in
Ref.~\cite{Detmold:2005gg} and expanded on in Refs.~\cite{Detmold:2018kwu,Detmold:2020lev,Detmold:2021uru,Detmold:2021qln,Detmold:2021plw}, which relates hadronic matrix elements directly computable in a Euclidean field theory to a heavy-quark operator product expansion (HOPE), where the non-perturbative information about the LCDA is encoded in its Mellin moments.  The moments are defined as
\begin{equation}
\braket{\xi^n}(\mu^2)=\frac{1}{2}\int_{-1}^1 d\xi\, \xi^n\phi_\pi(\xi,\mu^2)\,,
\end{equation}
and can be determined by fitting lattice data to the HOPE. The use of a fictitious heavy quark in the computation has the advantage that the heavy quark mass serves as a hard scale which suppresses higher-twist corrections, and can be varied to study their residual higher-twist effects present in the numerical data.

In Ref.~\cite{Detmold:2021qln}, the HOPE method was implemented to extract the second Mellin moment of the pion LCDA. The success of this approach motivates the current attempt to extend this formalism to obtain the fourth Mellin moment, $\braket{\xi^4}$, about which little is currently known. The only existing determination of the fourth Mellin moment from lattice QCD is $\braket{\xi^4}(\mu=2~\si{GeV})=0.124(11)(20)$ at a single lattice spacing of $a=0.076~\si{fm}$~\cite{Gao:2022vyh}. In this proceedings, progress towards the first continuum limit determination of the fourth Mellin moment of the pion LCDA is presented. The structure of this article is as follows: the HOPE method is briefly reviewed in Sec.~\ref{sec:formalism}, the numerical implementation is explained in Sec.~\ref{sec:numerical_implementation}, and finally the conclusions of this work are given in Sec.~\ref{sec:conclusion}. 

\section{The HOPE Method}
\label{sec:formalism}
The HOPE method has been previously discussed in Refs.~\cite{Detmold:2005gg,Detmold:2018kwu,Detmold:2020lev,Detmold:2021uru,Detmold:2021qln,Detmold:2021plw}. Here, only the main results will be restated.
The starting point of the approach for extracting moments of the pion LCDA is the hadronic matrix element
\begin{equation}
V^{[ \mu\nu ]} (q,p) =\int d^4z\, e^{iq\cdot z}\braket{0|T\{J_\Psi^\mu(z/2)J_\Psi^\nu(-z/2)\}|E^{(0)},\mathbf{p}}\,,
\end{equation}
where the current is given by $J_\Psi^\mu=\overline{\Psi}\gamma^\mu \gamma_5 \psi_l+\overline{\psi}_l\gamma^\mu \gamma_5 \Psi$ and $\Psi$ is the heavy quark field. The HOPE expression for this matrix element is~\cite{Detmold:2005gg,Detmold:2021uru}
\begin{equation}
\label{eq:Mellin_OPE_had_amp_target_mass}
V^{[ \mu\nu ]} (q,p) = - \frac{2 i \epsilon^{\mu\nu\rho\sigma}
 q_{\rho} p_{\sigma}}{\tilde{Q}^{2}}  f_{\pi} \sum_{n=0,{\mathrm{even}}}^{\infty}
 C_{W}^{(n)} (\tilde{Q}^{2},
 \mu, m_\Psi)  \braket{ \xi^{n} }  \left [ \frac{\zeta^{n} {\mathcal{C}}_{n}^{2}
(\eta)}{2^{n}(n+1)\tilde{Q}^{2}}\right ]\, ,
\end{equation}
where $m_\Psi$ is the heavy quark mass, $p^\mu$ and $q^\mu$ are the four-momenta of the pion and current, respectively. The scalar functions depend on the kinematic invariants
$\tilde{Q}^2=Q^2+m_\Psi^2$, $\zeta=\sqrt{p^2q^2}/\tilde{Q}^2$, and $\eta=p\cdot q/\sqrt{p^2q^2}$. The Wilson coefficients, $C_{W}^{(n)}$, have been computed in the $\overline{\text{MS}}$ scheme~\cite{Detmold:2021uru}, and thus the resulting heavy quark and Mellin moments are also to be understood in this scheme at the renormalization scale $\mu$. Performing a Fourier transform in the temporal direction, one obtains
\begin{equation}
\label{eq:fourier_transform}
R^{[\mu\nu]}(t,\mathbf{p},\mathbf{q})=\int \frac{dq_4}{(2\pi)}\, e^{-iq_4 t} V^{[ \mu\nu ]} (q,p)\,.
\end{equation}
This quantity can be determined from a ratio of two- and three-point correlators computed using LQCD.

\section{Numerical Implementation of HOPE Method}
\label{sec:numerical_implementation}
The quenched gauge fields used in this study were tuned to a constant physical volume of $L=1.92~\si{fm}$ and a constant pion mass of $m_\pi\sim 0.55~\si{GeV}$. Leading finite volume effects arise from the `around-the-world' pion contributions, which are small at this pion mass ($\exp(-m_\pi L)\approx 0.5\%$) and currently neglected in the analysis. Details on the lattice action, including the $\mathcal{O}(a)$ improvement obtained from the use of Wilson-clover fermions can be found in Ref.~\cite{Detmold:2021qln}.
Further details on the lattice ensembles and quark masses used are listed in Table~\ref{tab:lattice_details}.
The required two- and three-point functions were generated using the software package \textsc{Chroma} with the \textsc{QPhiX} inverters~\cite{chroma, qphix}. 

The starting point for a numerical implementation of the HOPE method is a calculation of certain two- and three-point correlators using LQCD. In particular, one computes the correlation functions 
\begin{equation}
 C_{ij}^{2} (t, \mathbf{p})
 = \int d^3 \mathbf{x} \, e^{i\mathbf{p}\cdot \mathbf{x}}  \braket{0|\mathcal{O}_i(t,\mathbf{x}) \mathcal{O}^\dagger_j (0, \mathbf{0}) |0}
\end{equation}
and
\begin{equation}
C_{i}^{3\mu\nu} (t_e, t_m; \mathbf{p}_e, \mathbf{p}_m)
 = \int d^3x_e \, d^3x_m\, e^{i\mathbf{p}_e\cdot \mathbf{x}_e}e^{i\mathbf{p}_m\cdot \mathbf{x}_m}\braket{0|{T}\left[ J_{l,\Psi}^\mu(\tau_e, \mathbf{x}_e) J_{l,\Psi}^\nu(\tau_m, \mathbf{x}_m) \mathcal{O}^\dagger_i(\mathbf{0}) \right] |0}\,,
\end{equation}
where $\mathcal{O}_i$ is a suitably chosen interpolating operator with the quantum numbers of a single pseudoscalar meson. The ambiguity in exactly how one chooses the interpolating operator can be exploited to suppress excited state contributions. Completeness of the energy eigenstates of the theory and translational invariance allows one to express this two-point correlator as
\begin{equation}
C_{ij}^{2}(t,\mathbf{p})=\sum_{n=0}^\infty \frac{Z_i^{(n)}(\mathbf{p})Z_j^{(n)*}(\mathbf{p})}{2E^{(n)}}e^{-E^{(n)}t}\,,
\label{eq:spectral_decomposition}
\end{equation}
where $Z_i^{(n)}(\mathbf{p})=\braket{0|\mathcal{O}_i(\mathbf{0},0)|E^{(n)},\mathbf{p}}$. It is desirable to choose an interpolating operator which has reduced overlap with excited state contributions so that a single exponential dominates Eq.~\eqref{eq:spectral_decomposition}. In this study, momentum smearing~\cite{Bali:2016lva} and the variational method~\cite{Michael:1982gb,Luscher:1990ck} are employed together to produce an optimized interpolating operator. Both of these techniques are discussed below.

For $0 \ll t \ll T$, the two-point correlator is saturated with the contribution of the lowest-lying hadronic state and can be written as
\begin{equation}
 C_{ij}^{(2)} (t, \mathbf{p}) = \frac{Z_i^{(0)}(\mathbf{p}) Z_j^{(0)*}(\mathbf{p})}{2E^{(0)}} e^{-E^{(0)} t} \bigg[1+A_{ij}e^{-\Delta E^{(1)} t}+\dots\bigg]\, ,
 \label{2-pt-func}
\end{equation}
which allows a determination of the overlap factor
$Z_i^{(0)}(\mathbf{p}) = \braket{0|\mathcal{O}_i | E^{(0)}, \mathbf{p} }$ and the pion ground state energy $E^{(0)}$. Excited state contributions are exponentially suppressed by the mass-gap $\Delta E^{(1)}=E^{(1)}-E^{(0)}$, and by the magnitude of the relative overlap factor, $A_{ij}$. Similarly, for $0 \ll t_e, t_m \ll T/2$, the three-point correlation function takes the form
\begin{equation}
 C_i^{(3)\mu\nu}(t_e, t_m; \mathbf{p}_e, \mathbf{p}_m) =
 R^{\mu\nu}(t; \mathbf{p}, \mathbf{q})
 \frac{Z_i^{(0)}(\mathbf{p})}{2E^{(0)}}
 e^{-E^{(1)}(t_e + t_m)/2}\bigg[1+B_{i} e^{-\Delta E^{(2)}(t_e + t_m)/2}+\dots \bigg] \, ,
\end{equation}
with $\mathbf{p} = \mathbf{p}_e + \mathbf{p}_m$, $t = t_e - t_m$ and $B_i$ is a relative overlap factor. Thus, one can divide the three-point correlator by the leading exponential behaviour and show that in the large Euclidean time limit, 
\begin{equation}\label{eq:r12}
\mathcal{R}(t)=\frac{2E^{(0)} C_i^{(3)\mu\nu}(t_e, t_m; \mathbf{p}_e, \mathbf{p}_m)}{Z_i^{(0)}(\mathbf{p}) e^{-E^{(0)}(t_e + t_m)/2}}\to R^{\mu\nu}(t; \mathbf{p}, \mathbf{q})\,.
\end{equation}
From this and Eqs.~\eqref{eq:Mellin_OPE_had_amp_target_mass} and \eqref{eq:fourier_transform}, one can extract the Mellin moments from a study of LQCD correlation functions.

\begin{table}
\centering
\setlength{\tabcolsep}{1em}
  \begin{tabular}{ c c c c c c c c }
  \hline\hline
$(L/a)^3 \times (T/a)$ & $\beta$ & $a$ (fm) & $N_\text{meas}$ & $\kappa_l$  & $\kappa_h$ & $m_\Psi^{\overline{\text{MS}}}$ (\si{GeV}) 
\\ \hline
\multirow{5}{*}{$24^3 \times 48$} & \multirow{5}{*}{6.10050} & \multirow{5}{*}{0.0813}  & \multirow{5}{*}{5000} & \multirow{5}{*}{0.134900}   
& 0.1300 & 1.2\\ 
& &  &  &  & 0.1250 & 1.7\\ 
& &  &  &  & 0.1200 & 2.0\\
& &  &  &  & 0.1160 & 2.3\\  
& &  &  &  & 0.1100 & 2.7\\ 
 \hline
\multirow{5}{*}{$32^3 \times 64$ } & \multirow{5}{*}{6.30168} & \multirow{5}{*}{0.0600}  & \multirow{5}{*}{5000} & \multirow{5}{*} {0.135154} 
& 0.125 &  1.7\\  
& &  &  &  &  0.1180 & 2.0\\
&  &  &  &  & 0.1130 & 3.1\\
 & &  &  &  &  0.1095 & 3.4\\
 \hline
 \multirow{3}{*}{$40^3 \times 80$ } & \multirow{3}{*}{6.43306} & \multirow{3}{*}{0.0502}  & \multirow{3}{*}{5000} & \multirow{3}{*}{0.135145} 
& 0.1270 &  2.0\\  
& &  &  &  &  0.1220 &  2.7\\
& &  &  &  &  0.1150 &  3.4\\
 \hline
 \multirow{5}{*}{$48^3 \times 96$} & \multirow{5}{*}{6.59773} & \multirow{5}{*}{0.0407}  & \multirow{5}{*}{2500} & \multirow{5}{*}{0.135027} & 0.1285 &  \multirow{5}{*}{-}\\  
& &  &  &  &  0.1244 &  \\
& &  &  &  & 0.1150 &   \\
& &  &  & & 0.1192 &  \\
& &  &  & & 0.1100 &  \\
   \hline
  \end{tabular}
    \caption{Details of the gauge field configurations and quark masses used in this study. These configurations were generated in Ref.~\cite{Detmold:2018zgk}. Heavy quark masses obtained from the fit of numerical data to the one-loop HOPE formula are also given. Measurements from the $48^3\times96$ ensemble was not included in the preliminary analysis reported here due to the lower statistics compared to the other lattice spacings. As a result, no values for the (fitted) heavy quark masses are given.}\label{tab:lattice_details}
\end{table}

\subsection{Operator Optimization}
In this section, an optimized interpolating operator for the pion is constructed by utilizing a combination of momentum smearing and the variational method. 
First introduced in Ref.~\cite{Bali:2016lva}, the technique of momentum smearing allows one to increase the operator overlap with hadronic states with finite three-momentum. As proposed in the original paper, this technique is paired with an implementation of gauge-invariant Gaussian smearing. The smeared operators may be written as
\begin{equation}
\psi^\text{sm}(\mathbf{x},x_4)=\int d^3y f(\mathbf{x}-\mathbf{y};U_\mu)\psi(\mathbf{y},x_4)\,.
\end{equation}
The function $f(\mathbf{x}-\mathbf{y};U_\mu)$ defines the specific smearing algorithm. The width of the Gaussian smearing was taken as $a\omega_\text{smear} = \{4.5, 6.0, 8.0, 9.0\}$ for $L/a = {24, 32, 40, 48}$, and a smearing momentum fraction of $\zeta = 0.8$ as
proposed in Ref.~\cite{Bali:2016lva} was employed.

By enlarging the set of operators, one can construct an improved variational estimate of the ground state operator. This study uses a two-operator set for the pion given by
\begin{align}
\mathcal{O}_1(\mathbf{x},t)&=\overline{\psi}^\text{sm}(\mathbf{x},t)\gamma_5 \psi^\text{sm}(\mathbf{x},t)\,,
\\
\mathcal{O}_2(\mathbf{x},t)&=\overline{\psi}^\text{sm}(\mathbf{x},t)\gamma_4\gamma_5 \psi^\text{sm}(\mathbf{x},t)\,.
\end{align}
While modern spectroscopy studies utilize larger sets of operators~\cite{Amarasinghe:2021lqa}, this study is only concerned with extracting ground state quantities, and thus the variational method is simply used to provide an operator with improved overlap to the ground state. Therefore the improvement gained by employing the two operators given above is sufficient for the current study. After computing the two-by-two correlator matrix defined by these operators, the generalized eigenvalue problem,
\begin{equation}
\sum_{j}C_{ij}(t,\mathbf{p})v_j^{(n)}(t,t_0,\mathbf{p})=\lambda^{(n)}(t,t_0)\sum_{j}C_{ij}(t_0,\mathbf{p})v_j^{(n)}(t,t_0,\mathbf{p})
\end{equation}
is solved. In principle, both the eigenvectors $v_j^{(n)}(t,t_0,\mathbf{p})$ and the corresponding eigenvalues $\lambda^{(n)}(t,t_0)$ are time dependent. An optimized interpolating operator for the ground state pion is thus given by
\begin{equation}
\mathcal{O}^{(0)}(\mathbf{x},t)=\sum_i v_i^{(0)}(t_\text{ref},t_0,\mathbf{p}) \mathcal{O}_i(\mathbf{x},t)
\end{equation}
In this study, both $t_0$ and $t_\text{ref}$ are fixed to ensure that matrix elements computed with this operator are linear combinations of LQCD correlation functions and thus exhibit a simple spectral representation. This optimized interpolating operator is used in calculations of both the two-point and three-point correlation functions.

\subsection{Extracting the Mellin Moments}
\begin{figure}
    \centering
    \includegraphics[scale=0.45]{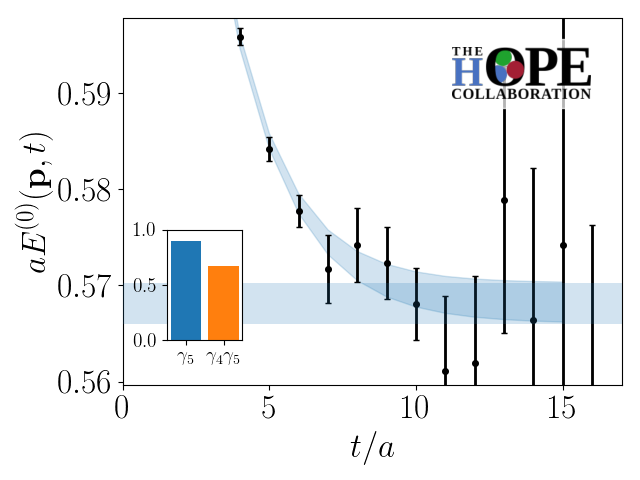}
    \includegraphics[scale=0.45]{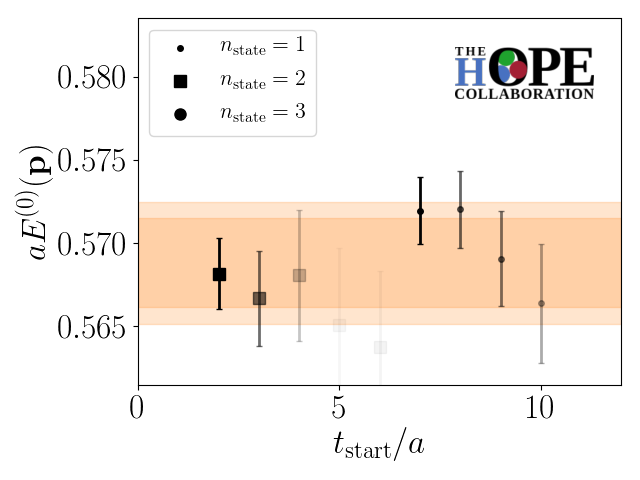}
    \caption{Two-point spectroscopy is used to extract a variational estimate of the ground-state energy. The left-hand plot shows the effective mass for the optimized interpolating operator. The optimized ground state correlator is fit to an $n$-state model for each $t_\text{start}/a$. $n$ is chosen according to an AIC criteria. The shaded curve shows the highest weighted fit. The resulting fits are combined in a weighted average.
    The right-hand plot shows the fitted values of the ground state energy as a function of $t_\text{start}/a$. More heavily weighted fits are colored darker. The shaded band is the result.}
    \label{fig:two-point_spectroscopy}
\end{figure}

In order to implement the HOPE method, a standard spectroscopic analysis of the two-point correlator data obtained from the optimized interpolating operator is required. Fits to this data are performed following the algorithm outlined in Ref.~\cite{Amarasinghe:2021lqa}. In particular, fits are performed for a range of $t_\text{start}$ for fixed $t_\text{stop}$. A weighted average of these fits is then performed using the weighting procedure outlined in Ref.~\cite{NPLQCD:2020ozd}. An example of this procedure is shown in Fig.~\ref{fig:two-point_spectroscopy}. 

\begin{figure}
    \centering
    \includegraphics[scale=0.5]{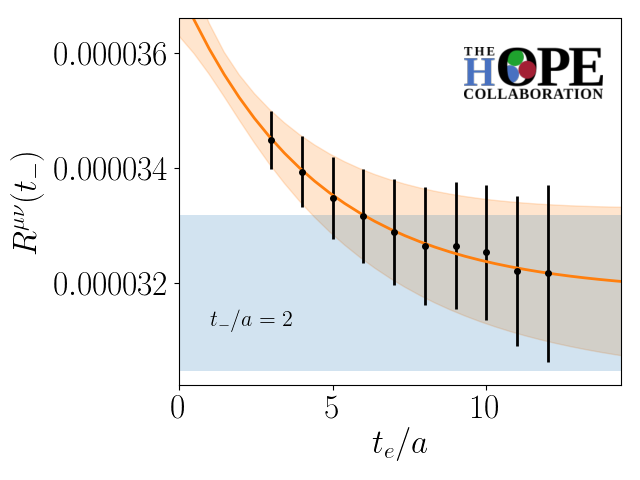}
    \caption{Study of excited state in ratio constructed from Eq.~\eqref{eq:r12}. From this analysis, one can see excited state contributions are small for $t_e/a>6$.}
    \label{fig:excited_state}
\end{figure}

Having extracted a variational estimate of the required ground-state energy, it is now possible to construct the ratio in Eq.~\eqref{eq:r12}. By computing this matrix element for a range of $t_e/a=\{3,4,5,6,7,8,9,10,11,12\}$ on the $24^3\times 48$ ensemble, one can analyse the excited state contamination in the matrix element $R^{\mu\nu}(t)$. This is demonstrated in Fig.~\ref{fig:excited_state}. From this study it is possible to see that excited state contamination is small for $t_e/a>6$. Since these three-point correlation functions require the use of the sequential source method, additional $t_e$ values require an approximately linear increase in the total computational cost. In this analysis, data was generated using $t_e/a=7,8$ corresponding to $t_e=\{0.56~\si{fm},0.65~\si{fm}\}$ in physical units. The same distances in physical units were used for $t_e$ on the other gauge field ensembles.

\begin{figure}
    \centering
    \includegraphics[scale=0.45]{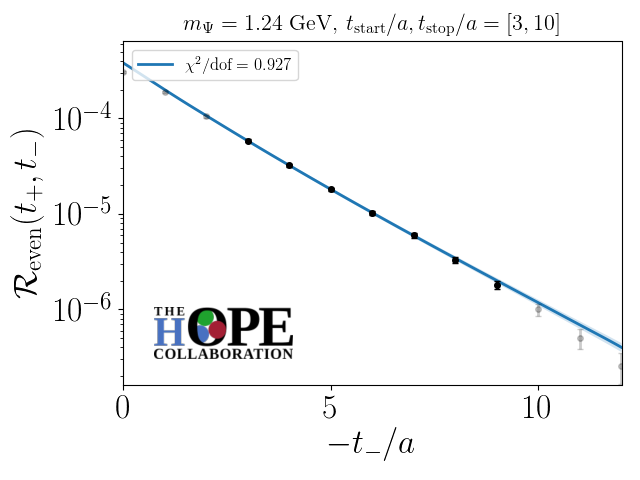}
    \includegraphics[scale=0.45]{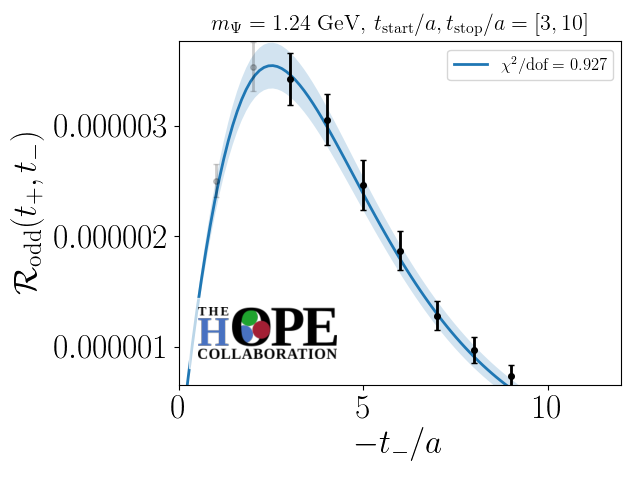}
    \caption{$t$-even and $t$-odd components of ratio obtained from computing Eq.~\eqref{eq:r12}. Data are fitted to a truncated, Fourier transformed version of Eq.~\eqref{eq:Mellin_OPE_had_amp_target_mass}. }
    \label{fig:ratio}
\end{figure}

The resulting numerical calculation of Eq.~\eqref{eq:r12} for $L/a=24$ with a heavy quark mass of $m_\Psi\sim 1.2~\si{GeV}$ is shown in Fig.~\eqref{fig:ratio}. By fitting the Fourier transform of Eq.~\ref{eq:Mellin_OPE_had_amp_target_mass} truncated to include the second and fourth Mellin moments to this numerical data, information about the Mellin moments of the pion LCDA can be obtained. Since this fit is perfomed at finite lattice spacing, it is important to note that these results contain residual lattice artifacts, and may be contaminated with higher-twist contributions due to the truncation of the HOPE to the leading twist-two contribution only. In the next section, a combined continuum, twist-two extrapolation of the fitted lattice data will be described.

\subsection{Continuum, Twist-Two Extrapolation}
\begin{figure}
    \centering
    \includegraphics[scale=0.45]{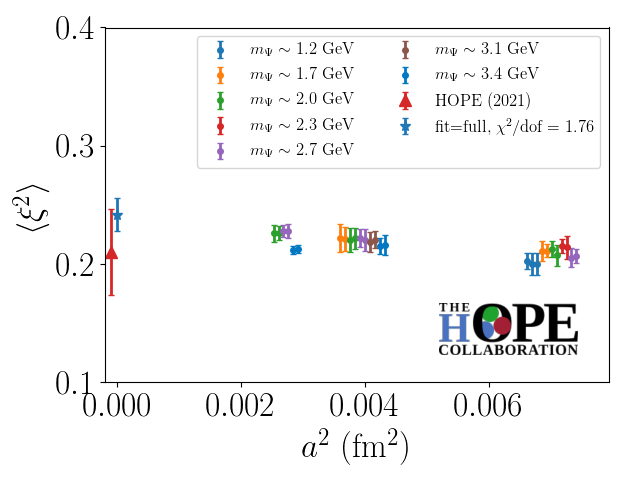}
    \includegraphics[scale=0.45]{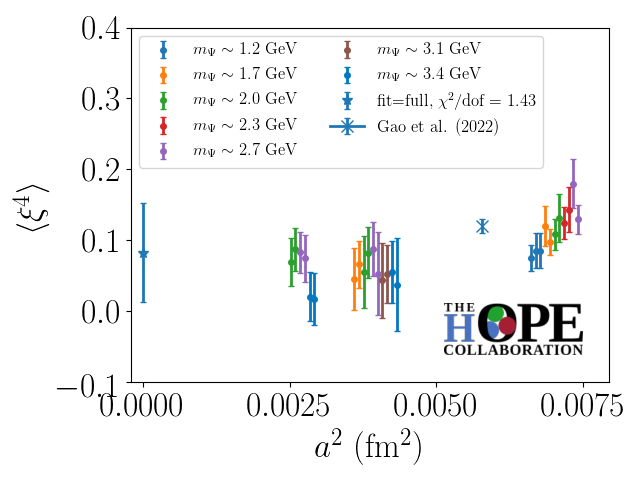}
    \caption{Preliminary continuum, twist-two extrapolation of fitted Mellin moments. The analysis was performed using data from the two $t_e$ values. The extrapolated values are $\braket{\xi^2}(\mu^2=4~\si{GeV}^2)=0.245\pm0.014$ and $\braket{\xi^4}(\mu^2=4~\si{GeV}^2)=0.075\pm0.070$. The second Mellin moment is in good agreement with the previous HOPE determination of this quantitiy using the same lattice ensembles.}
    \label{fig:continuum_twist2_extrapolation}
\end{figure}

A combined continuum, twist-two extrapolation must be performed to connect the numerical data obtained in the previous section with the pion LCDA moments. 
The parameterization is~\cite{Detmold:2021qln}
\begin{equation}
\braket{\xi^n}(\mu^2,a,m_\Psi)=\braket{\xi^n}(\mu^2)+\frac{A_n(\mu^2)}{m_\Psi}+B_n(\mu^2)a^2+C_n(\mu^2)a^2m_\Psi +D_n(\mu^2)a^2m_\Psi^2 \, ,
\end{equation}
where $A$, $B$, $C$ and $D$ are dimensionless parameters fit to the numerical data, and the renormalization scale ($\mu^2$) dependence is given explicitly. Global fits to all ensembles and heavy quark masses are performed independently for the second and fourth moment data. The resulting combined continuum, twist-two extrapolated moments are shown in Fig.~\ref{fig:continuum_twist2_extrapolation}. 
This analysis finds
\begin{align}
\braket{\xi^2}(\mu^2=4~\si{GeV}^2)=0.245\pm0.014\,,&
\\
\braket{\xi^4}(\mu^2=4~\si{GeV}^2)=0.075\pm0.070\,,&
\end{align}
in the quenched approximation with $m_\pi=550~\si{MeV}$, where the quoted errors are purely statistical. The result for the second Mellin moment is in good agreement with the previous HOPE determination of the quantity using the same gauge field samples~\cite{Detmold:2021qln}. While the statistical uncertainty for the fourth Mellin moment is large, it is important to note that this preliminary analysis constitutes the first continuum limit determination of this quantity from LQCD. Data taking for this project is still in progress, with the aim of reducing the statistical errors in this combined continuum, twist-two extrapolation.

\section{Conclusion}
\label{sec:conclusion}
In this proceedings, progress towards a determination of the fourth Mellin moment of the pion LCDA using the HOPE method is presented. The current analysis used numerical results computed at three lattice spacings. Data on a fourth lattice spacing closer to the continuum limit is being computed. A combination of momentum smearing and the variational method is used to improve the overlap of the interpolating operator with the ground state pion and reduce excited state contamination. The resulting optimized interpolating operator is used to compute the hadronic matrix element appearing in the HOPE method. After fitting the HOPE to the numerical data, the resulting fit parameters are extrapolated to the continuum, twist-two limit, where they can be identified with the Mellin moments of the pion LCDA. As a result of this extrapolation, this report presents the first preliminary continuum limit determination of the fourth Mellin moment from LQCD. 

\acknowledgments
The authors thank ASRock Rack Inc.~for their support of the construction of an Intel Knights Landing cluster at National Yang Ming Chiao Tung University, where the numerical calculations were performed.  Help from Balint Joo in tuning Chroma is acknowledged. We thank M. Endres for providing the ensembles of gauge field configurations used in this work. The authors thankfully acknowledge the computer resources at MareNostrum and the technical support provided by BSC (RES-FI-2023-1-0030).
WD and AVG are supported by U.S.~Department of Energy under grant Contract Numbers DE-SC0011090 and DE-SC0023116. WD is further supported by 
the National Science Foundation under Cooperative Agreement PHY-2019786 (The NSF AI Institute for Artificial Intelligence and Fundamental Interactions, http://iaifi.org/).
AVG was additionally supported by the resources of the Fermi National Accelerator Laboratory (Fermilab), a U.S. Department of Energy, Office of Science, Office of High Energy Physics HEP User Facility.
CJDL is supported by the Taiwanese NSTC grant number 112-2112-M-A49 -021 -MY3.
RJP has been supported by project PID2020-118758GB-I00, financed by the Spanish MCIN/ AEI/10.13039/501100011033/.
YZ is supported by the U.S. Department of Energy, Office of Science, Office of Nuclear Physics through Contract No.~DE-AC02-06CH11357

\addcontentsline{toc}{chapter}{Bibliography} 
\bibliographystyle{jhep.bst} 
\bibliography{bibliography} 

\end{document}